\documentclass[prb,showpacs,showkeys,preprintnumbers,amsmath,amssymb,twocolumn]{revtex4-1}
\usepackage[T1]{fontenc} 
\usepackage{makeidx} 
\usepackage{graphicx} 
\usepackage{dcolumn} 
\usepackage{array} 
\usepackage{amssymb} 
\usepackage{amsmath}
\usepackage{textcomp}
\usepackage{rotating}
\usepackage{wasysym}
\usepackage{multirow}
\usepackage{subfigure}
\usepackage{color}
\usepackage{eucal}
\usepackage{mathrsfs}
\usepackage{units}
\usepackage{rotating}
\usepackage[all]{xy}
\usepackage{float} 
\usepackage{amsmath} 
\usepackage{amsfonts} 
\usepackage{bm} 
\usepackage[amssymb]{SIunits}

\begin{document}

\title{Magneto-oscillations on specific heat of graphene monolayer}

\author{Z.Z. Alisultanov}\email{zaur0102@gmail.com}
\affiliation{Amirkhanov Institute of Physics Russian Academy of Sciences, Dagestan Science Centre, Russia, 367003, Makhachkala, Yagarskogo str., 94}
\affiliation{Prokhorov General Physics Institute Russian Academy of Sciences, Russia, 119991, Moscow, Vavilov Str., 38}
\affiliation{Dagestan States University, Russia, 367000, Makhachkala, Gadzhiyev Str., 43-a}
\author{M.S. Reis}\email{marior@if.uff.br}\affiliation{Instituto de F\'{i}sica, Universidade Federal Fluminense, Av. Gal. Milton Tavares de Souza s/n, 24210-346, Niter\'{o}i-RJ, Brasil}
\keywords{de Haas-van Alphen effect, graphenes, specific heat, density of states}

\date{\today}

\begin{abstract}
Measurement of magnetic oscillations on thermodynamic quantities (like magnetization and specific heat), is one of the experimental methods to access the density of states of electronic systems. In the present paper we therefore theoretically explore the oscillatory phenomena on the specific heat of graphenes considering gapped and gapless cases in a quantized magnetic field. Further situations is also considered, as the influence of impurities, Coulomb interaction and phonons. We could then map the magnetic oscillations on the specific heat of graphenes under these constraints and the obtained results are a good starting point and guide for further experimental works.
\end{abstract}

\maketitle

\section{Introduction}
The band structure of monolayer graphenes shows that it is a gapless semiconductor with a linear energy spectrum near the so-called Dirac points\cite{um} and, as one of many consequences,   Landau levels (LLs) are non equidistant; and therefore unusual behaviors upon a magnetic field change is observed\cite{dois}. In addition, the distance between Landau levels is very large (for example, the energy gap between the first two LLs in a magnetic field of 10 T is more than 1000 K). This leads to giant magneto-optical\cite{tres,quatro} and thermo-magnetic effects\cite{cinco,alisultanov2015oscillating,alisultanov2014oscillating,alisultanov2015oscillating,paixao2014oscillating,reis2014diamagnetic,reis2014step,reis2013oscillating,reis2013electrocaloric,reis2013influence,reis2012oscillating,reis2012oscillating,reis2011oscillating}, as well as to an unusual quantum Hall effect, which can be observed even at room temperature\cite{um,seis,sete,oito}. 

Due to the valence of carbon ions and the gapless energy spectrum, the electronic density of states (DOS) of graphenes vanishes at the Fermi energy and therefore any further knowledge on this issue is useful for the scientific community. In this direction, experimental access to the DOS could be obtained by measurements of magnetic oscillations of the heat capacity under a quantizing magnetic field\cite{nove,dez}. 

Thus, the present effort theoretically explores the oscillatory behavior of the specific heat of monolayer graphenes under a constant area, considering the gapless and gapped cases; and, in addition, the influence of impurities, Coulomb interaction and phonons. The obtained results are therefore a good start point for further experimental studies.

\section{Magneto-oscillations of electronic specific heat}

The thermodynamic potential of the two-dimensional electron gas in a quantized magnetic field can written as\cite{alisultanov2014oscillating}:
\begin{equation}\label{omegarkwelewf}
\Omega(T)=-\frac{2eHk_BT\mathcal{S}}{\pi \hbar c}\sum_n\ln\left[1+\exp\left(\frac{\mu-\epsilon_n}{k_BT}\right)\right]
\end{equation}
where $H$ is the magnetic field, $e$ represents the electron charge, $c=3\times10^{10}$ cm/s the light speed, $\mathcal{S}$ stands for the graphene area and $\mu>0$ represents the chemical potential. From now on, we will use the quasi-classical approach, based on the quantization conditions of Lifshitz-Onsager\cite{onze}. Considering the plane of the electronic system along $xy$ directions and an ortogonal magnetic field (along $z$ axis), this approach starts considering a quantized area enclosed by an electron trajectory in momentum space:
\begin{equation}
A(\epsilon)=\frac{2\pi \hbar eH}{c}(j+\gamma_\pm)
\end{equation}
Above,
\begin{equation}
\gamma_\pm=\gamma\pm\frac{m^*}{2m},
\end{equation}
where (i) $\gamma$ is a constant that defines the energy of the zeroth Landau level (it assumes $1/2$ for non-relativistic gas and zero for graphenes), and (ii) $m^*/2m$ is the Zeeman splitting. In addition:
\begin{equation}
m^*=\frac{1}{2\pi}\frac{dA(\epsilon)}{d\epsilon},
\end{equation}
is the electron cyclotron mass and $m$ the electron mass. Finally, $j$ is an integer and $\epsilon(p)=\epsilon$.

For instance, considering free electrons on graphenes we would have $A(\epsilon)=\pi \epsilon^2/v_f^2$ and then $m^*=\epsilon/v_F^2$. For graphenes, the values of chemical potential $0<|\mu|<1$ eV are of most interest, since the energy spectrum is linear for these values\cite{um}; and thus, from this interval, $m^*$ ranges from zero up to $0.1m$. These intervals 
have an interesting consequence: the absolute value of the Zeeman splitting then ranges from zero up to 0.05, and therefore, since for the present case we are considering $|\mu|<1$ eV, we can thus neglect the Zeeman splitting. 

It is well known that the thermodynamic potential of electrons in a quantizing magnetic field consists of two components\cite{dezesseis,dezessete}: a zero-field contribution $\Omega_0(T)$ and a field dependent term $\Omega_H^{osc}(T)$. Thus, the total potential reads as:
\begin{equation}
\Omega(T)=\Omega_0(T)+\Omega_H^{osc}(T)
\end{equation}
where\cite{dezessete}
\begin{equation}
\Omega_0(T)=-\frac{\mathcal{S}}{(\pi \hbar)^2}\int_0^\infty A(\epsilon)f(\epsilon-\mu)d\epsilon
\end{equation}
and\cite{dezessete}
\begin{equation}
\Omega_H^{osc}(T)=\frac{2m^*\omega_ck_BT\mathcal{S}}{\pi \hbar}\sum_{k=1}^\infty \frac{1}{k}\frac
{\cos(kn\pi)}
{\sinh(x_k)}
\end{equation}
Above, $f(\epsilon-\mu)$ represents the Fermi-Dirac distribution function,
\begin{equation}\label{npifer}
n\pi=\frac{A(\mu)c}{\hbar eH},
\end{equation}
\begin{equation}
x_k=k\frac{2\pi^2k_BT}{\hbar\omega_c},
\end{equation}
and
\begin{equation}
\omega_c=\frac{eH}{m^*c}
\end{equation}
where this last means the cyclotron frequency.

From the above, the entropy of the system can be easily accessed from $S=-(\partial \Omega/\partial T)_\mu$, and reads therefore as:
\begin{equation}\label{eq_six}
S_0(T)=k_B\frac{\mathcal{S}}{4\pi^2\hbar^2}\frac{1}{k_B^2T^2}\int_0^\infty\frac{\epsilon-\mu}{\cosh^2\left[\frac{\epsilon-\mu}{2k_BT}\right]}A(\epsilon)d\epsilon
\end{equation}
and
\begin{equation}
S_H^{osc}(T)=k_B\frac{\pi m^*k_BT\mathcal{S}}{\hbar^2}\sum_{k=1}^\infty\frac{L(x_k)\cos(kn\pi)}{\sinh x_k}
\end{equation}
where
\begin{equation}
L(x)=\coth x-\frac{1}{x}
\end{equation}
is the Langevin equation. If $\epsilon_F\gg k_BT$ then integrand in equation \ref{eq_six} is appreciably different from zero only in the vicinity of  $\epsilon=\mu\approx \epsilon_F$; and thus we can expand the function $A(\epsilon)$ near $\epsilon=\epsilon_F$. This assumption leads to:
\begin{equation}
A(\epsilon)\approx A(\epsilon_F)+2\pi m^*_F(\epsilon-\epsilon_F)
\end{equation}
where $m^*_F=m^*(\epsilon_F)$. Equation \ref{eq_six} can then be rewritten as:
\begin{equation}
S_{0}(T)\approx k_B\frac{2\pi m^*_F\mathcal{S} k_BT}{3\hbar^2}
\end{equation}

However, in a quantized magnetic field, the chemical potential is an oscillating function of the magnetic field; but the total number of electrons $N$ must remains a constant and therefore we can consider:
\begin{equation}
N=-\left(\frac{\partial \Omega}{\partial\mu}\right)_{H,T}=\frac{\mathcal{S}A(\epsilon_F)}{(\pi\hbar)^2}-\left(\frac{\partial\Omega^{osc}_H}{\partial \mu}\right)_{H,T}
\end{equation}
where the first term was obtained considering the $\epsilon_F\gg k_BT$ limit on equation \ref{eq_six}. Considering also this limit for the second term we then obtain (see reference \onlinecite{cinco}):
\begin{equation}\label{f3frwf}
\frac{A(\epsilon_F)c}{\hbar eH}=\alpha-\arctan\left[\frac{\sin(2\alpha)}{1+\cos(2\alpha)}\right]
\end{equation}
where
\begin{equation}
\alpha=\frac{\pi^2\hbar c}{eH}\mathcal{N}
\end{equation}
and $\mathcal{N}=N/\mathcal{S}\approx 10^{12}$ cm$^{-2}$ is the electronic concentration. Note the left side of equation \ref{f3frwf} stands for $n\pi$, on equation \ref{npifer}. For gapless monolayer graphene it reads as\cite{dois}
\begin{equation}\label{gaplesswfwce}
A(\epsilon_F)=\frac{\pi\epsilon_F^2}{v_F^2};
\end{equation}
while for a gapped monolayer graphene\cite{treze,dezenove}
\begin{equation}\label{gappedffw}
A(\epsilon_F)=\frac{\pi}{v_F^2}(\epsilon_F^2-\Delta^2)
\end{equation}
and, finally, for a 2D non-relativistic electrons gas it reads as\cite{onze}
\begin{equation}
A(\epsilon_F)=2\pi m\epsilon_F
\end{equation}

Finally, the specific heat (at constant graphene area $\mathcal{S}$) can be determined from the entropy above, considering $c_a=T(\partial S/\partial T)_a$. Thus, the total area specific heat read as 
\begin{equation}\label{crkwfwfwojk}
c_a(T)=\frac{\pi m^*_Fk_BT}{\hbar^2}\left\{\frac{2}{3}+\sum_{k=1}^\infty\mathcal{F}(x_k)\frac{\cos(kn\pi)}
{\sinh x_k}\right\}k_B
\end{equation}
where
\begin{equation}\label{funcaore}
\mathcal{F}(x_k)=2\coth x_k-x_k\left(\frac{1}{\sinh^2 x_k}+\coth^2 x_k\right)
\end{equation}
It should be noted that we consider only the intraband contribution to the specific heat, i.e., we considered only temperatures $k_BT\ll\hbar\omega_c$.

\begin{figure}
\begin{center}
\subfigure{\includegraphics[angle=270,width=9cm]{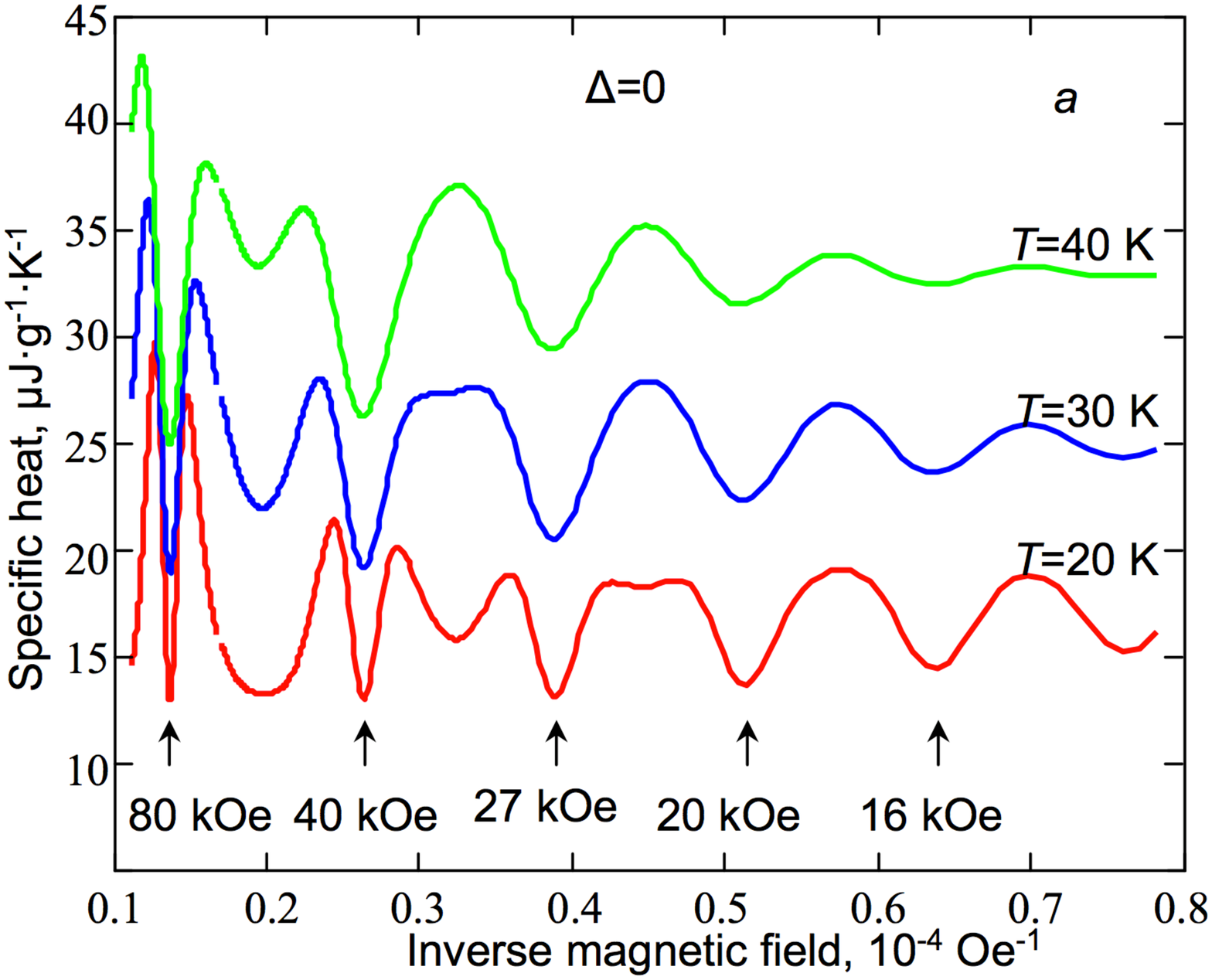}}
\subfigure{\includegraphics[angle=270,width=9cm]{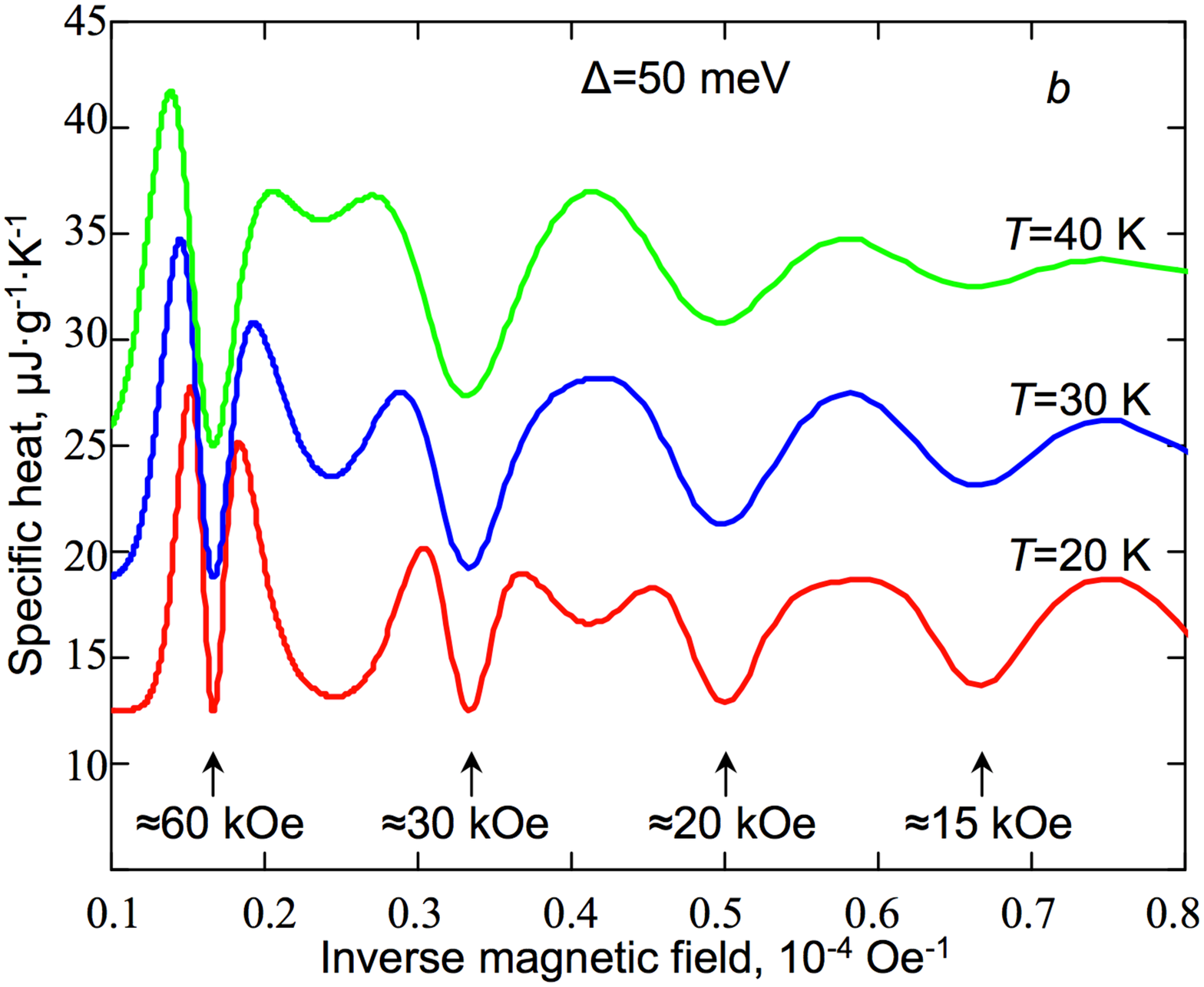}}
\end{center}
\caption{Oscillatory area specific heat for (a) gapless and (b) gapped graphenes.}\label{umfig}
\end{figure}

Since there are a cosine function with argument depending on the inverse magnetic field $H$, it is expected that the specific heat oscillates when represented as a function of $1/H$; and this effect is caused by quantum oscillations of the density of states. See figure \ref{umfig} for a further understanding, where two cases are depicted: (a) gapless and (b) gapped graphenes. This oscillations can be seen as follows: magnetic field values in which a certain LL is almost fully filled (and the immediately above LL is still completely empty) promotes a sharp change on the density of states and, consequently, on the total energy and specific heat. A small increasing of the magnetic field thus fills the LL and then the specific heat reaches to a minimum - and this fact can be physically understood, since completed LLs does not exchange energy (inter-level transition is not allowed, due to the low temperature under consideration; and intra-level transition is also not allowed, since the level is completely filled). 

It is also important to stress a changing on the oscillation period:
\begin{equation}
\Delta\left(\frac{1}{H}\right)\approx\frac{2\pi\hbar e}{A(\epsilon_F)c},
\end{equation}
comparing the gapped and gapless cases. See equations \ref{gaplesswfwce} and \ref{gappedffw}, as well as figure \ref{umfig}.

\section{Landau level broadening}

Real samples have defects and impurities and therefore the scattering by these leads to a broadening of the Landau levels (known as Dingle broadening). The thermodynamic potential considering this broadening of LLs can be written as:
\begin{equation}
\Omega(T)=-k_BT\int_{-\infty}^\infty\rho(\epsilon)\ln\left[1+\exp\left(\frac{\mu-\epsilon}{k_BT}\right)\right]d\epsilon
\end{equation}
In the absence of broadening of the Landau levels, the density of states is written as:
\begin{equation}
\rho(\epsilon)=\frac{2eH\mathcal{S}}{\pi \hbar c}\sum_n\delta(\epsilon-\epsilon_n)
\end{equation}
leading therefore to equation \ref{omegarkwelewf}; however, to consider the the broadening effect, the density of states reads as:
\begin{equation}
\rho(\epsilon)=\frac{2eH\mathcal{S}}{\pi \hbar c}\sum_n\mathcal{D}(\epsilon-\epsilon_n)
\end{equation}
where
\begin{equation}
\mathcal{D}(\epsilon)=\frac{1}{\pi}\frac{\Gamma}{\epsilon^2+\Gamma^2}
\end{equation}
is the Lorentzian approach and 
\begin{equation}
\mathcal{D}(\epsilon)=\frac{1}{\Gamma\sqrt{\pi}}e^{-\epsilon^2/\Gamma^2}
\end{equation}
represents the Gaussian one (the present work considers only the Lorentzian approach). Above, $\Gamma$ is the scattering energy. A detailed study of magnetic oscillations in graphenes in the presence of impurities, considering a Lorentzian distribution, is detailed discussed in reference \onlinecite{dezessete}; and, from this work, the thermodynamic potential could be obtained:
\begin{equation}
\Omega_{H,\Gamma}^{osc}(T)=\frac{2m^*\omega_ck_BT\mathcal{S}}{\pi\hbar}\sum_{k=1}^\infty\frac{1}{k}\exp\left[-2\pi k\frac{\Gamma}{\hbar\omega_c}\right]
\frac{\cos(kn\pi)}{\sinh(x_k)}
\end{equation}

Considering $\epsilon_F\gg\Gamma$, the zero field contribution $\Omega_{0,\Gamma}(T)$ can be neglected; and, on the other hand, $\Gamma\gg\hbar\omega_c$, completely damps the oscillations of the field dependent contribution (above equation). Thus, the total area specific heat, taking the broadening of the LLs and $\Gamma\ll\hbar\omega_c$ into account, reads as:
\begin{align}
c_{a,\Gamma}(T)=&\frac{\pi m^*k_BT}{\hbar^2}\left\{\frac{2}{3}+\right.\\\nonumber
&\left.\sum_{k=1}^\infty\mathcal{F}(x_k)\exp\left[-2\pi k\frac{\Gamma}{\hbar\omega_c}\right]\frac{\cos(kn\pi)}{\sinh x_k}\right\}k_B
\end{align}
where $\mathcal{F}(x_k)$ is given by equation \ref{funcaore}.

\begin{figure}
\begin{center}
\subfigure{\includegraphics[angle=270,width=9cm]{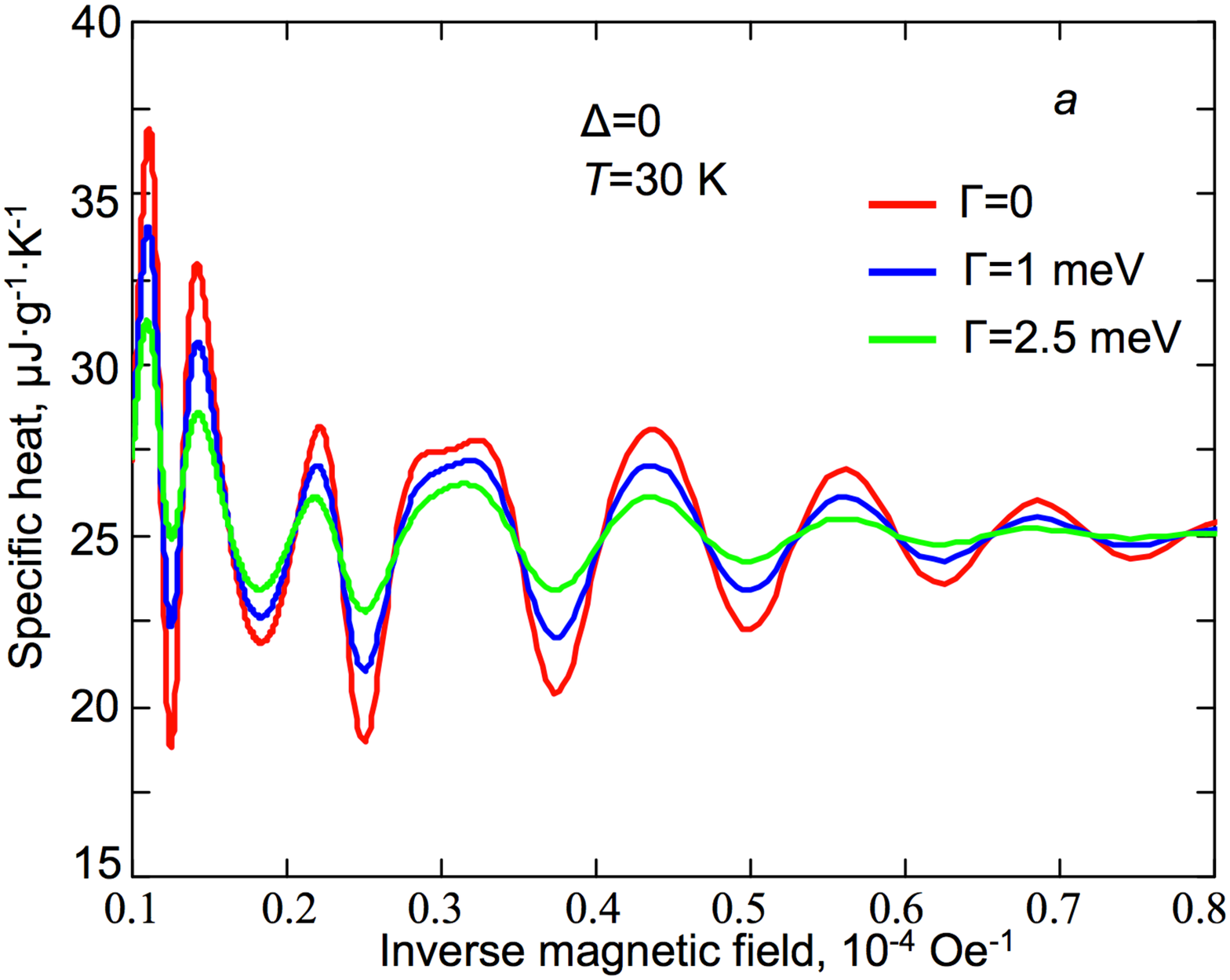}}
\subfigure{\includegraphics[angle=270,width=9cm]{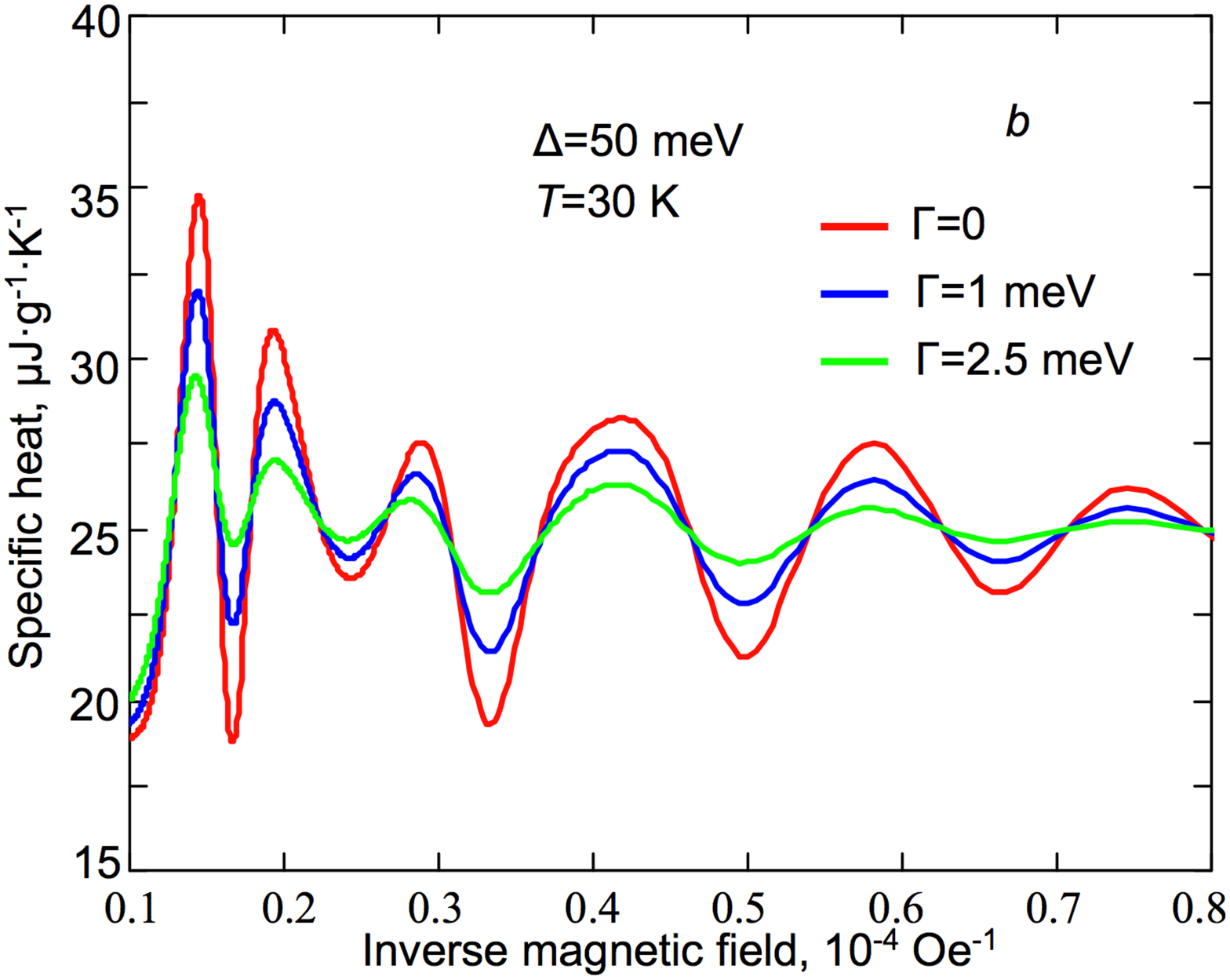}}
\end{center}
\caption{Influence of impurities on the oscillatory area specific heat for (a) gapless and (b) gapped graphenes. These values of $\Gamma$ are those that satisfy $\Gamma\ll\hbar\omega_c$. \label{doisfig}}
\end{figure}

Indeed, increasing the scattering parameter $\Gamma$, the oscillations are suppressed, as can be clearly seen on figure \ref{doisfig}(a), for gapless graphenes, and \ref{doisfig}(b), for gapped case. These results are of easy understand, since impurities make broader the LLs, decreasing the density of states.

\section{Influence of Coulomb interaction}

Coulomb interaction is a very important problem in graphene science, since this interaction is not screened due to the two-dimensional character of the system. To consider this interactions into the quasi-classical approach, we need to change the conditions of quantization. Strictly speaking, nowadays, the problem of the influence of the Coulomb interaction on the magnetic oscillations is not solved; however, in some studies\cite{dezoito,dezenove,vinte} this interaction has been introduced into the function $A(\epsilon)$ without changing the quantization condition. In this paper, we will also use this simple approach.

Thus, the electronic spectrum of graphene renormalized by the Coulomb interaction reads as\cite{vinteum,vintedois}:
\begin{equation}
\epsilon(p)=\nu_bv_Fp\left[1+g\ln\left(\frac{p_0}{p}\right)\right]
\end{equation}
where $g=e^2/8\pi\hbar v_F\kappa$, $\kappa$ is the dielectric constant (substrate plus graphene - for SiO$_2$ it is close to 3.9, see reference \onlinecite{hwang2007dielectric}), $p_0=5\times10^{-20}$ erg.s/cm\cite{vintedois} and $\nu_b$ assumes `+' for the conduction band and `-' for the valence band. It is possible to show\cite{dezoito}, that the electron trajectory in momentum space, considering the first order expansion in order to small values of $g$, is a circle of radius 
\begin{equation}
a(\epsilon)=\frac{\epsilon}{v_F}\left[1-g\ln\left(\frac{v_Fp_0}{\epsilon}\right)\right];
\end{equation}
and therefore
\begin{equation}
A(\epsilon_F)=\frac{\pi\epsilon_F^2}{v_F^2}\left[1-g\ln\left(\frac{v_Fp_0}{\epsilon_F}\right)\right]^2
\end{equation}

\begin{figure}{b!}
\center
\includegraphics[angle=270,width=9cm]{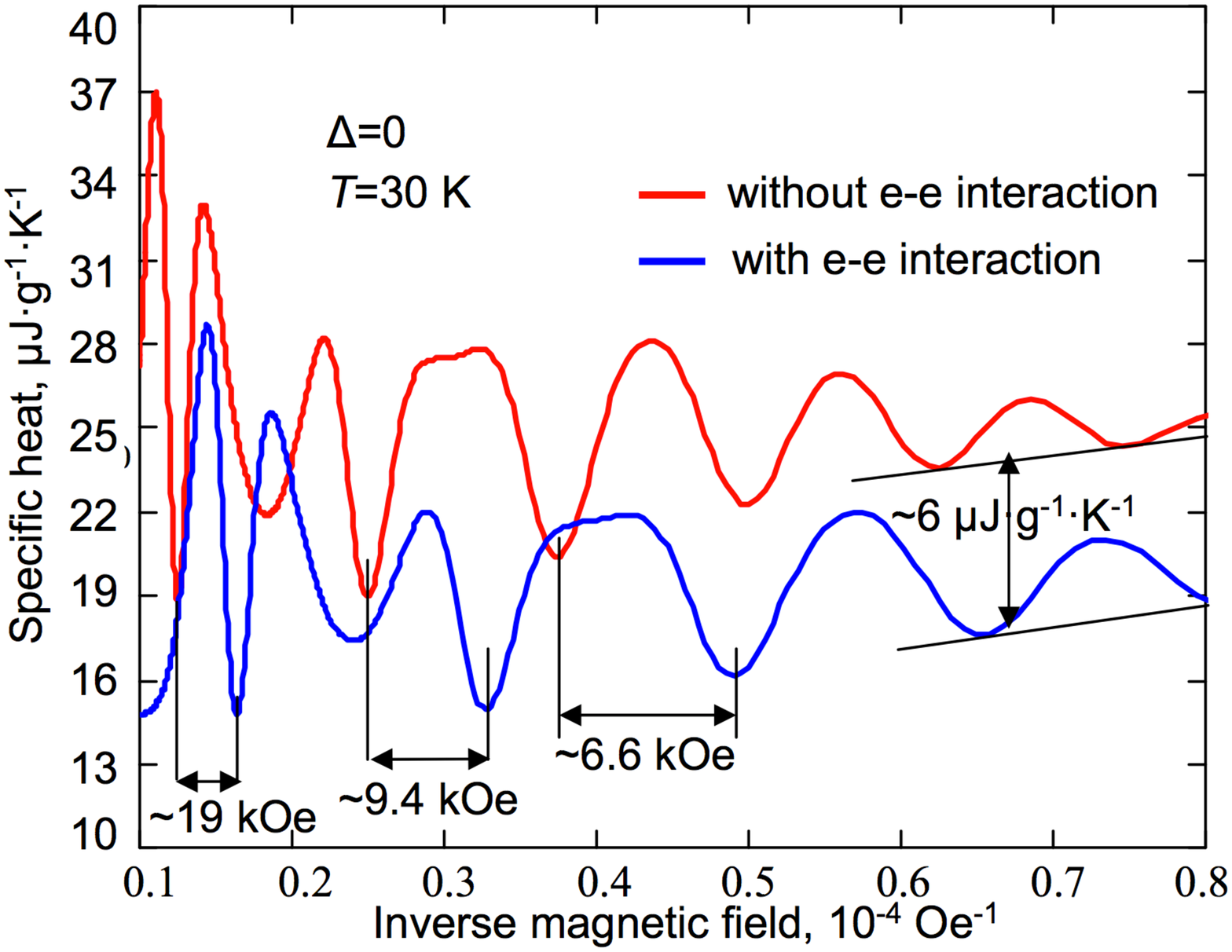}
\caption{Influence of Coulomb interaction on the oscillatory area specific heat for gapless graphenes. \label{tresfig}}
\end{figure}

The above quantity, for a gapless graphene, can then be placed into equation \ref{npifer} and the specific heat of equation \ref{crkwfwfwojk} revisited. This result is depicted on figure \ref{tresfig}, where the expected oscillations in the presence and absence of the electron-electron interaction can be compared. Note the Coulomb interaction leads to a decreasing of the specific heat, as well as to an increasing and shifting of the oscillations. Apparently, this is due to the fact that the Coulomb interaction favors a decreasing of the chemical potential\cite{dezoito}. 

\section{Phonons contribution}

The total specific heat of the system, in addition to the electronic term, also contains a phonon contribution; that, on its turn, is usually more pronounced than the electronic one. As done above, to access this contribution, we need the thermodynamic potential, that, for this case, reads as\cite{landau1980statistical}: 
\begin{equation}
\Omega_{ph}(T)=\frac{\mathcal{S}k_BT}{2\pi \vartheta_{ph}^2}\int_0^\infty\omega\ln\left[1-\exp\left(-\frac{\hbar\omega}{k_BT}\right)\right]d\omega
\end{equation}
where $\omega$ is the phonon frequency and $\vartheta_{ph}$ is the phonon velocity (in graphene $\vartheta_{ph}\approx 10^6$ cm/s)\cite{vintetres}. Introducing a new variable $y=\hbar\omega/k_BT$  and integrating by parts, we obtain:
\begin{equation}
\Omega_{ph}(T)=-\frac{\mathcal{S}k_B^3T^3}{4\pi \vartheta_{ph}^2\hbar^2}\int_0^\infty\frac{y^2}{e^y-1}dy
\end{equation}
that, after a simple integration reads as:
\begin{equation}
\Omega_{ph}(T)=-\frac{3\mathcal{S}k_B^3T^3}{8\pi \vartheta_{ph}^2\hbar^2}\zeta(3)
\end{equation}
where  $\zeta(z)$ is the Riemann zeta function. Then, the phonon entropy is a straightforward result:
\begin{equation}
S_{ph}(T)=\frac{9\mathcal{S}k_B^2T^2}{8\pi \vartheta_{ph}^2\hbar^2}\zeta(3)k_B
\end{equation}
and, consequently, the total area specific heat:
\begin{align}
c_a(T)=&\frac{\pi m(\mu)k_BT}{\hbar^2}\times\\\nonumber
&\left\{\frac{2}{3}+\sum_{k=1}^\infty\mathcal{F}(x_k)\frac{\cos(kn\pi)}{\sinh x_k}+\frac{9k_BT}{4\pi^2\vartheta_{ph}^2m(\mu)}\zeta(3)\right\}k_B
\end{align}

\begin{figure}
\center
\includegraphics[angle=270,width=9cm]{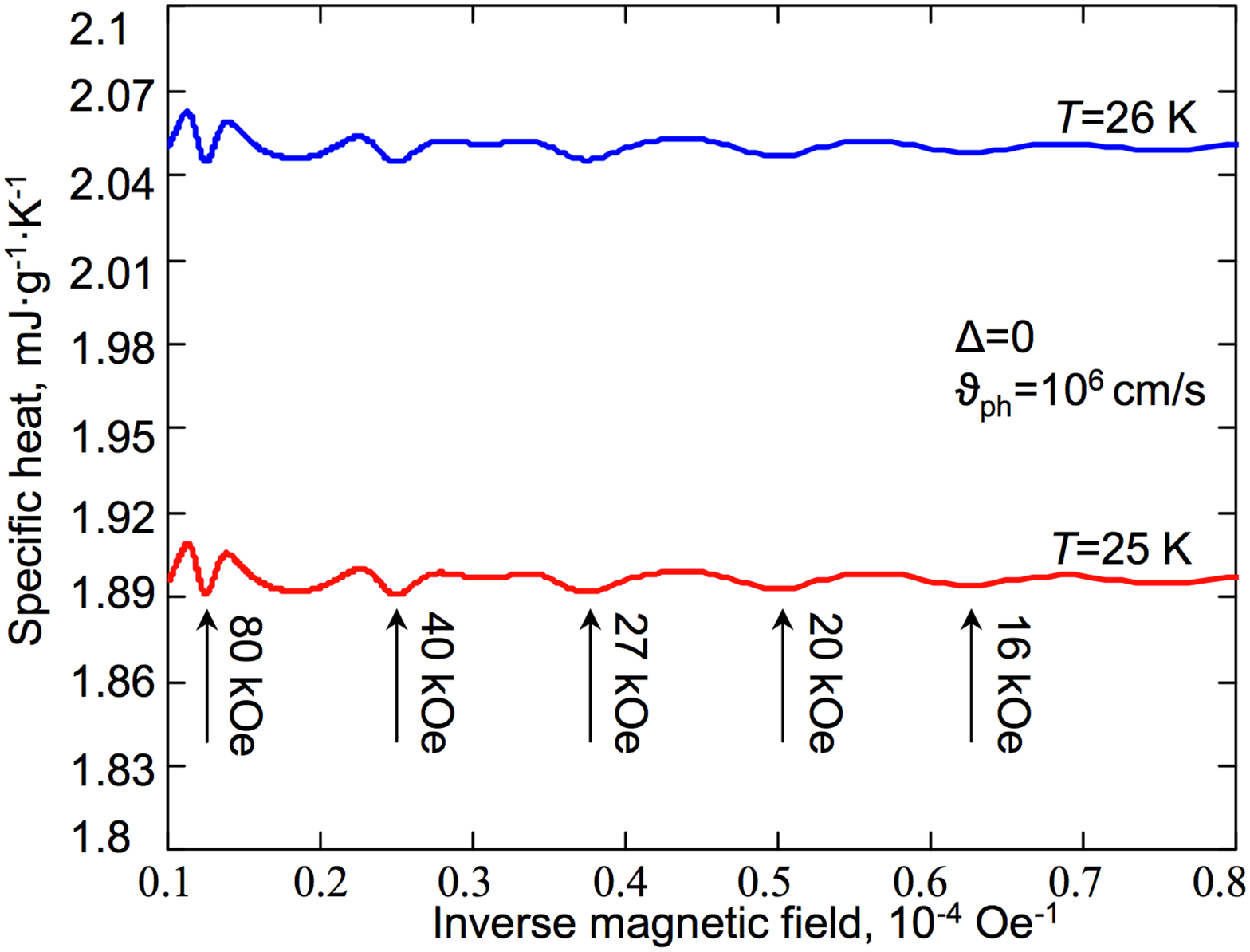}
\caption{Electronic plus phonon area specific heat for gapless graphenes. Note this case is three orders of magnitude bigger than the other cases.
\label{quatrofig}}
\end{figure}

This final result can then be seen in figure \ref{quatrofig}.  Note the phonon part of the specific heat is much greater than the electronic one (oscillations remain, but the total value is three orders of magnitude bigger). Quantitatively, the ratio of the phonon and electronic parts of the specific heat agrees with results of reference \onlinecite{vintetres}. 

\section{Conclusions}

The present work evaluates the electronic specific heat under constant area for gapped and gapless graphenes. Due to the crossing of the Landau level with the Fermi energy of the system, oscillations are found, in analogy to the de Haas-van Alphen effect, found on the magnetization of these materials\cite{zhang2010modulation}, and also on the magnetocaloric potentials\cite{alisultanov2015oscillating,alisultanov2014oscillating,alisultanov2015oscillating,paixao2014oscillating,reis2014diamagnetic,reis2014step,reis2013oscillating,reis2013electrocaloric,reis2013influence,reis2012oscillating,reis2012oscillating,reis2011oscillating}. Impurities, Coulomb interaction and phonons are also considered and deeply evaluated. Considering these oscillations are useful for an experimental access to the Fermi surface of materials\cite{coldea2008fermi}, the present results are therefore a good start point to use specific heat as a probe of the Fermi surface of the studied cases.

\begin{acknowledgements}
ZZA declare that this work was supported by grant RFBR number 15-02-03311a, President Grant MK-4471.2015.2 and project number 3.1262.2014  of the Ministry of Education and Science of the Russian Federation in the field of scientific research. ZZA is also sincerely grateful to Dmitry Zimin Foundation "Dynasty" for financial support. MSR thanks FAPERJ, CAPES, CNPq and PROPPI-UFF for financial support. 
\end{acknowledgements}

\end{document}